\documentclass[preprint,aps,amssymb,showpacs,superscriptaddress,nofootinbib]{revtex4}
\usepackage{graphicx}

\newcommand{\Case}[2]{{\textstyle \frac{#1}{#2}}}
\newcommand{\lP}{\ell_{\mathrm P}}

\begin{document}

\preprint{IMSc/2007/03/2}

\title{Singularity Resolution in Isotropic Loop Quantum Cosmology: Recent Developments}

\author{Ghanashyam Date}
\email{shyam@imsc.res.in}
\affiliation{The Institute of Mathematical Sciences,
CIT Campus, Chennai-600 113, INDIA}


\begin{abstract}

Since the past Iagrg meeting in December 2004, new developments in loop
quantum cosmology have taken place, especially with regards to the
resolution of the Big Bang singularity in the isotropic models.  The
singularity resolution issue has been discussed in terms of physical
quantities (expectation values of Dirac observables) and there is also
an ``improved'' quantization of the Hamiltonian constraint. These
developments are briefly discussed. 

This is an expanded version of the review talk given at the
24$^{\mathrm{th}}$ IAGRG meeting in February 2007.

\end{abstract}

\maketitle

\section{Cosmology, quantum cosmology, loop quantum cosmology}

Our current understanding of large scale properties of the universe is
summarised by the so called $\Lambda-$CDM Big Bang model -- homogeneous
and isotropic, spatially flat space-time geometry with a positive
cosmological constant and cold dark matter. Impressive as it is, the
model is based on an Einsteinian description of space-time geometry
which has the Big Bang singularity.  The existence of cosmological
singularities is in fact much more general. There are homogeneous but
anisotropic solution space-times which are singular and even in the
inhomogeneous context there is a general solution which is singular
\cite{BKL}. The singularity theorems give a very general argument for
the existence of initial singularity for an everywhere expanding
universe with normal matter content, the singularity being characterized
as incompleteness of causal geodesic in the past.  Secondly, in
conjunction with an inflationary scenario, one imagines the origin of
the smaller scale structure to be attributed to quantum mechanical
fluctuations of matter and geometry.  On account of both the features, a
role for the quantum nature of matter {\em and} geometry is indicated.

Quantum mechanical models for cosmological context were in fact
constructed, albeit formally. For the homogeneous and isotropic sector,
the geometry is described by just the scale factor and the extrinsic
curvature of the homogeneous spatial slices. In the gravitational
sector, a quantum mechanical wave function is a function of the scale
factor i.e. a function on the (mini-) superspace of gravity. The scale
factor being positive, the minisuperspace has a boundary and wave
functions need to satisfy a suitable boundary condition. Furthermore,
the singularity was not resolved in that the Wheeler-De Witt equation (or
the Hamiltonian constraint) which is a differential equation with
respect to the scale factor, had singular coefficient due to the matter
density diverging near the boundary.  Thus quantization per se does not
necessarily give a satisfactory replacement of the Big Bang singularity.

Meanwhile, over the past 20 years, a background independent,
non-perturbative quantum theory of gravity is being constructed starting
from a (gauge-) connection formulation of classical general relativity
\cite{LQGRev}.  The background independence provided strong constraints
on the construction of the quantum theory already at the kinematical
level (i.e. before imposition of the constraints) and in particular
revealed a discrete and non-commutative nature of quantum (three
dimensional Riemannian) geometry. The full theory is still quite
unwieldy.  Martin Bojowald took to step of restricting to homogeneous
geometries and quantizing such models in a {\em loopy way}. Being
inherited from the connection formulation, the geometry is described in
terms of densitized triad which for the homogeneous and isotropic
context is described by $p \sim \mathrm{sgn}(p) a^2$ which can also take
negative values (encoding the orientation of the triad). This means that
the classical singularity (at $p = 0$) now lies in the interior of the
superspace. Classically, the singularity prevents any relation between
the two regions of positive and negative values of $p$. Quantum
mechanically however, the wave functions in these two regions, could be
related. One question that becomes relevant in a quantum theory is that
if a wave function, specified for one orientation and satisfying the
Hamiltonian constraint, can be unambiguously extended to the other
orientation while continuing to satisfy the Hamiltonian constraint.
Second main implication of loop quantization is the necessity of using
holonomies -- exponentials of connection variable $c$ -- as well defined
operators.  This makes the Hamiltonian constraint a difference equation
on the one hand and also requires an indirect definition for inverse
triad (and inverse volume) operators entering in the definition of the
matter Hamiltonian or densities and pressures.  Quite interestingly, the
Hamiltonian constraint equation turns out to be non-singular (i.e.
deterministic) and in the effective classical approximation suggests
interesting phenomenological implication quite naturally. These two
features in fact made LQC an attractive field.

We will briefly summarise the results prior to 2005 and then turn to
more recent developments. An extensive review of LQC is available in
\cite{LQCRev}. For simplicity and definiteness, we will focus on the
spatially flat isotropic models.

\section{Summary of pre 2005 LQC}

{\em Classical model:} Using coordinates adapted to the spatially
homogeneous slicing of the space-time, the metric and the extrinsic
curvature are given by, 
\begin{equation} 
ds^2 := - dt^2 + a^2(t)\left\{(dr^2 + r^2d\Omega^2\right\} \ .  
\end{equation}
Starting from the usual Einstein-Hilbert action and scalar matter for
definiteness, one can get to the Hamiltonian as,
\begin{eqnarray}
S & := & \int dt \int_{\mathrm{cell}} dx^3 \sqrt{|det
g_{\mu\nu}|}\left\{ \frac{R(g)}{16 \pi G} + \frac{1}{2}\dot{\phi}^2 -
V(\phi) \right\} \nonumber\\
& = & V_0\int dt \left\{\frac{3}{8 \pi G}(- a\dot{a}^2) + \frac{1}{2}a^3
\dot{\phi}^2 - V(\phi) a^3 \right\} \\
p_a & = & - \frac{3 V_0}{4 \pi G} a \dot{a} ~ ~ , ~ ~ p_{\phi} ~ = ~ V_0
a^3 \dot{\phi} ~ ~,~ ~ V_0 ~ := ~ \int_{\mathrm{cell}} d^3x ~ ;
\nonumber\\
H (a, p_a, \phi, p_{\phi}) & = & H_{\mathrm{grav}} + H_{\mathrm{matter}}
\nonumber \\
& = & \left[- \frac{2 \pi G}{3} \frac{p_a^2}{V_0 a} \right] +
\left[\frac{1}{2} \frac{p_{\phi}^2}{a^3 V_0} + a^3 V_0 V(\phi)\right] \\
& = & \left(\frac{3 V_0 a^3}{8 \pi G}\right)\left[ -
\frac{\dot{a}^2}{a^2} + \left(\frac{8 \pi G}{3}\right)
\left(\frac{H_{\mathrm{matter}}}{ V_0 a^3}\right)\right]
\end{eqnarray}
Thus, $H = 0 \leftrightarrow $ Friedmann Equation. For the spatially
flat model, one has to choose a fiducial cell whose fiducial volume is
denoted by $V_0$.

In the connection formulation, instead of the metric one uses the
densitized triad i.e. instead of the scale factor $a$ one has
$\tilde{p}, |\tilde{p}| := a^2/4$ while the connection variable is
related to the extrinsic curvature as: $\tilde{c} := \gamma \dot{a}/2$
(the spin connections is absent for the flat model). Their Poisson
bracket is given by $\{\tilde{c}, \tilde{p}\} = (8\pi G \gamma)/(3
V_0)$. The arbitrary fiducial volume can be absorbed away by defining $c
:= V_0^{1/3} \tilde{c}, ~ p := V_0^{2/3}\tilde{p}$. Here, $\gamma$ is
the Barbero-Immirzi parameter which is dimensionless and is determined
from the Black hole entropy computations to be approximately $0.23$
\cite{BHEntropy}. From now on we put $8\pi G := \kappa$. The classical
Hamiltonian is then given by,
\begin{equation}\label{ClassHam}
H ~ = ~ \left[- \frac{3}{\kappa}\left( \gamma^{-2}c^2 \sqrt{|p|}\right)
\right] + \left[\frac{1}{2}|p|^{-3/2} p_{\phi}^2 + |p|^{3/2}
V(\phi)\right] \ .
\end{equation}
For future comparison, we now take the potential for the scalar field,
$V(\phi)$ to be zero as well.

One can obtain the Hamilton's equations of motion and solve them easily.
On the constrained surface ($H = 0$), eliminating $c$ in favour of $p$
and $ p_{\phi}$, one has,
\begin{eqnarray}
c ~ = ~ \pm \gamma \sqrt{\frac{\kappa}{6}} \frac{|p_{\phi}|}{|p|} ~ & ,
& ~ \dot{p} ~ = ~ \pm \sqrt{\frac{\kappa}{6}} |p_{\phi}||p|^{-1/2} \ .
\nonumber \\
\dot{\phi} ~ = ~ p_{\phi} |p|^{-3/2}~ & , & ~ \dot{p_{\phi}} ~ = ~ 0\ ,
\\
\frac{d p}{d \phi} ~ = ~ \pm \sqrt{\frac{2\kappa}{3}} |p| ~ &
\Rightarrow & ~  
p(\phi) ~ = ~ p_* e^{\pm \sqrt{\frac{2\kappa}{3}}( \phi - \phi_* )}
\label{ClassRelationalSoln}
\end{eqnarray}
Since $\phi$ is a monotonic function of the synchronous time $t$, it can
be taken as a new ``time'' variable. The solution is determined by
$p(\phi)$ which is (i) independent of the constant $p_{\phi}$ and (ii)
passes through $p = 0$ as $\phi \to \pm \infty$ (expanding/contracting
solutions).  It is immediate that, along these curves, $p(\phi)$, the
energy density and the extrinsic curvature diverge as $p \to 0$.
Furthermore, the divergence of the density implies that $\phi(t)$ is
{\em incomplete} i.e. $t$ ranges over a semi-infinite interval as $\phi$
ranges over the full real line  \footnote{For the FRW metric, integral
curves of $\partial_t$ are time-like geodesics and hence incompleteness
with respect to $t$ is synonymous with geodesic incompleteness.}.  Thus
a singularity is signalled by a solution $p(\phi)$ passing through $p =
0$ in {\em finite} synchronous time (or equivalently by the density
diverging somewhere along the solution).  A natural way to ensure that
all solutions are non-singular is to ensure that either of the two terms
in the Hamiltonian constraint are {\em bounded}. Question is: {\em does
a quantum theory replace the Big Bang singularity by something
non-singular?}.

There are at least two ways to explore this question. One can imagine
computing corrections to the Hamiltonian constraint such that individual
terms in the effective constraint are bounded. Alternatively and more
satisfactorily, one should be able to define suitable observables whose
expectation values will generate the analogue of $p(\phi)$ curves along
which physical quantities such as energy density, remain bounded.  The
former method was used pre-2005 because it could be used for more
general models (non-zero potential, anisotropy etc). The latter has been
elaborated in 2006, for the special case of vanishing potential. Both
methods imply that classical singularity is resolved in LQC but not in
Wheeler-De Witt quantum cosmology. We will first discuss the issue in
terms of effective Hamiltonian because it is easier and then discuss it
in terms of the expectation values.

In the standard Schrodinger quantization, one can introduce wave
functions of $p, \phi$ and quantize the Hamiltonian operator by $c \to
i\hbar \kappa\gamma/3 \partial_p ~,~ p_{\phi} \to -i \hbar
\partial_{\phi}$, in equation (\ref{ClassHam}). With a choice of
operator ordering, $\hat{H}\Psi(p, \phi) = 0$ leads to the
Wheeler-De Witt partial differential equation which has singular
coefficients.

The background independent quantization of Loop Quantum Gravity however
suggest a different quantization of the isotropic model. One should look
for a Hilbert space on which only exponentials of $c$ (holonomies of the
connection) are well defined operators and not $\hat{c}$. Such a Hilbert
space consists of almost periodic functions of $c$ which implies that
the triad operator has every real number as a proper eigenvalue:
$\hat{p}|\mu\rangle := \Case{1}{6}\gamma \lP^2 \mu|\mu\rangle, \forall
\mu \in \mathbb{R}~ , \lP^2 := \kappa\hbar$. This has a major
implication: {\em inverses of positive powers of triad operators do not
exist} \cite{ABL}. These have to be defined by using alternative
classical expressions and promoting them to quantum operators. This can
be done with at least one parameter worth of freedom, eg.
\begin{equation}
|p|^{-1} ~ = ~ \left[ \frac{3}{8\pi G\gamma l}\{c, |p|^l\}\right]^{1/(1
-l)} ~,~ l \in (0, 1)\ .  
\end{equation}
Only positive powers of $|p|$ appear now. However, this still cannot be
used for quantization since there is no $\hat{c}$ operator. One must use
holonomies: $h_j(c) ~ := ~ e^{\mu_0 c \Lambda^i\tau_i}\ ,$ where
$\tau_i$ are anti-hermitian generators of $SU(2)$ in the $j^{th}$
representation satisfying $\mathrm{Tr}_j (\tau_i \tau_j) = - \Case{1}{3}
j (j + 1) (2j + 1) \delta_{ij}$, $\Lambda^i$ is a unit vector specifying
a direction in the Lie algebra of $SU(2)$ and $\mu_0$ is the coordinate
length of the loop used in defining the holonomy. Using the holonomies,
\begin{eqnarray} 
|p|^{-1} & = & (8 \pi G \mu_0\gamma l)^{\frac{1}{l - 1}} \left[
\frac{3}{j(j + 1)(2j + 1)} \mathrm{Tr}_j \Lambda\cdot\tau \
h_j\{h_j^{-1}, |p|^l\}\right]^{\frac{1}{1 - l}} \ ,
\end{eqnarray}
which can be promoted to an operator. Two parameters, $\mu_0 \in
\mathbb{R}$ and $j \in \mathbb{N}/2$, have crept in and we have a three
parameter family of inverse triad operators. The definitions are:
\begin{eqnarray}
\widehat{|p|^{-1}_{(jl)}} |\mu\rangle & = &
\left(\frac{2j\mu_0}{6}\gamma\lP^2\right)^{-1} (F_{l}(q))^{\frac{1}{1
-l}} |\mu\rangle ~ ~ , ~ ~ q := \frac{\mu}{2\mu_0j} ~ := ~
\frac{p}{2jp_0}~ ~,\nonumber \\
F_l(q) & := & \frac{3}{2l}\left[ ~ ~ \frac{1}{l + 2}\left\{ (q + 1)^{l +
2} - |q - 1|^{l + 2}\right\} \right. \nonumber \\ & & \hspace{0.7cm}
\left. - \frac{1}{l + 1} q \left\{ (q + 1)^{l + 1} - \mathrm{sgn}(q -1)
|q - 1|^{l + 1}\right\} ~ ~ \right] \\
F_l( q \gg 1 ) & \approx & \left[q^{-1}\right]^{1 - l} \ , \nonumber \\
F_l( q \approx 0 ) & \approx & \left[\frac{3q}{l + 1}\right] \ . 
\end{eqnarray}
All these operators obviously commute with $\hat{p}$ and their
eigenvalues are bounded above. This implies that the matter densities
(and also intrinsic curvatures for more general homogeneous models),
remain bounded at the classically singular region. Most of the
phenomenological novelties are consequences of this particular feature
predominantly anchored in the matter sector.  We have thus two scales:
$p_0 := \Case{1}{6}\mu_0\lP^2$ and $2jp_0 := \Case{1}{6}\mu_0 (2j)
\lP^2$. The regime $|p| \ll p_0$ is termed the {\em deep quantum
regime}, $p \gg 2jp_0$ is termed the {\em classical regime} and $p_0
\lesssim |p| \lesssim 2jp_0$ is termed the {\em semiclassical regime}.
The modifications due to the inverse triad defined above are strong in
the semiclassical and the deep quantum regimes. For $j = 1/2$ the
semiclassical regime is absent.  Note that such scales are not available
for the Schrodinger quantization.

The necessity of using holonomies also imparts a non-trivial structure
for the gravitational Hamiltonian. The expression obtained is:
\begin{eqnarray}
H_{\mathrm{grav}} & = & - \frac{4}{8 \pi G \gamma^3
\mu_0^3} \sum_{ijk} \epsilon^{ijk}\mathrm{Tr}\left(h_i h_j
h_i^{-1}h_j^{-1}h_k\{h_k^{-1}, V\}\right)
\end{eqnarray}
In the above, we have used $j = 1/2$ representation for the holonomies
and $V$ denotes the volume function.  In the limit $\mu_0 \to 0$ one
gets back the classical expression.

While promoting this expression to operators, there is a choice of
factor ordering involved and many are possible. We will present two
choices of ordering: the {\em non-symmetric} one which keeps the
holonomies on the left as used in the existing choice for the full
theory, and the particular {\em symmetric} one used in \cite{APSOne}. 
\begin{eqnarray}
\hat{H}^{\mathrm{non-sym}}_{\mathrm{grav}} & = & \frac{24 i }{\gamma^3
\mu_0^3 \lP^2} \mathrm{sin}^2 \mu_0c \left( \mathrm{sin}\frac{\mu_0c}{2}
\hat{V} \mathrm{cos}\frac{\mu_0c}{2} - \mathrm{cos}\frac{\mu_0c}{2}
\hat{V} \mathrm{sin}\frac{\mu_0c}{2} \right) \\
\hat{H}^{\mathrm{sym}}_{\mathrm{grav}} & = & \frac{24 i
(\mathrm{sgn}(p))}{\gamma^3 \mu_0^3 \lP^2} \mathrm{sin} \mu_0c \left(
\mathrm{sin}\frac{\mu_0c}{2} \hat{V} \mathrm{cos}\frac{\mu_0c}{2} -
\mathrm{cos}\frac{\mu_0c}{2} \hat{V} \mathrm{sin}\frac{\mu_0c}{2}
\right) \mathrm{sin} \mu_0c \end{eqnarray}
At the quantum level, $\mu_0$ cannot be taken to zero since $\hat{c}$
operator does not exist. The action of the Hamiltonian operators on
$|\mu\rangle$ is obtained as,
\begin{eqnarray}
\hat{H}^{\mathrm{non-sym}}_{\mathrm{grav}}|\mu\rangle & = &
\frac{3}{\mu_0^3\gamma^3\lP^2}\left(V_{\mu + \mu_0} - V_{\mu -
\mu_0}\right) \left( |\mu + 4 \mu_0\rangle - 2 | \mu\rangle + |\mu -
4\mu_0\rangle\right) \\
\hat{H}^{\mathrm{sym}}_{\mathrm{grav}}|\mu\rangle & = &
\frac{3}{\mu_0^3\gamma^3\lP^2}\left[ \left|V_{\mu + 3\mu_0} - V_{\mu +
\mu_0}\right||\mu + 4 \mu_0\rangle + \left|V_{\mu - \mu_0} - V_{\mu -
3\mu_0}\right||\mu - 4 \mu_0\rangle \right. \nonumber \\ & & \left.
\hspace{1.5cm} - \left\{\left|V_{\mu + 3\mu_0} - V_{\mu + \mu_0}\right|
+ \left|V_{\mu - \mu_0} - V_{\mu - 3\mu_0}\right|\right\}|\mu\rangle
\right]
\end{eqnarray}
where $V_{\mu} := (\Case{1}{6}\gamma \lP^2 |\mu|)^{3/2}$ denotes the
eigenvalue of $\hat{V}$. Denoting quantum wave function by $\Psi(\mu,
\phi)$ the Wheeler-De Witt equation now becomes a {\em difference}
equation. For the non-symmetric one we get,
$$
A(\mu + 4 \mu_0) \psi(\mu + 4\mu_0, \phi) - 2 A(\mu) \psi(\mu, \phi) +
A(\mu - 4 \mu_0) \psi(\mu - 4\mu_0, \phi) 
$$
\begin{equation} 
~ = ~ -\frac{2 \kappa}{3}\mu_0^3 \gamma^3\lP^2 H_{matter}(\mu)\psi(\mu,
\phi)
\end{equation} 
where, $A(\mu) := V_{\mu + \mu_0} - V_{\mu - \mu_0}$ and vanishes for
$\mu = 0$.

For the symmetric operator one gets,
$$
f_+(\mu) \psi(\mu + 4\mu_0, \phi) + f_0(\mu) \psi(\mu, \phi) +
f_-(\mu) \psi(\mu - 4\mu_0, \phi) 
$$
\begin{equation} 
~ = ~ -\frac{2 \kappa}{3}\mu_0^3 \gamma^3\lP^2 H_{matter}(\mu)\psi(\mu,
\phi) \hspace{3.0cm} \mbox{where,}
\end{equation} 
$$
f_+(\mu) ~ := ~ \left| V_{\mu + 3\mu_0} - V_{\mu + \mu_0} \right| ~ , ~ 
f_-(\mu) ~ := ~ f_+(\mu - 4 \mu_0) ~, ~ 
f_0 ~ := ~  - f_+(\mu) - f_-(\mu) \ . 
$$
Notice that $f_+(-2\mu_0) = 0 = f_-(2\mu_0)$, but $f_0(\mu)$ is never
zero. The absolute values have entered due to the sgn($p$) factor. 

These are effectively second order difference equations and the
$\Psi(\mu, \phi)$ are determined by specifying $\Psi$ for two
consecutive values of $\mu$ eg for $ \mu = \hat{\mu} + 4 \mu_0 ,
\hat{\mu}$ for some $\hat{\mu}$. Since the highest (lowest) order
coefficients vanishes for some $\mu$, then the corresponding component
$\Psi(\mu, \phi)$ is undetermined by the equation. Potentially this
could introduce an arbitrariness in extending the $\Psi$ specified by
data in the classical regime (eg $\mu \gg 2j $) to the negative $\mu$.

For the non-symmetric case, the highest (lowest) $A$ coefficients vanish
for their argument equal to zero thus leaving the corresponding $\psi$
component undetermined. However, this undetermined component is
decoupled from the others. Thus apart from admitting the trivial
solution $\psi(\mu, \phi) := \Phi(\phi)\delta_{\mu, 0}, ~ \forall \mu$,
all other non-trivial solutions are completely determined by giving two
consecutive components: $\psi(\hat{\mu}, \phi), \psi(\hat{\mu} + 4\mu_0,
\phi)$.

For the symmetric case, due to these properties of the $f_{\pm,0}(\mu)$,
it looks as if the difference equation is {\em non-deterministic} if
$\mu = 2\mu_0 + 4\mu_0 n, n \in \mathbb{Z}$. This is because for $\mu =
-2\mu_0$, $\psi(2\mu_0, \phi)$ is undetermined by the lower order
$\psi$'s and this coefficient enters in the determination of
$\psi(2\mu_0, \phi)$.  However, the symmetric operator also commutes
with the parity operator: $(\Pi\psi)(\mu, \phi) := \psi(-\mu, \phi)$.
Consequently, $\psi(2\mu_0, \phi)$ is determined by $\psi(-2\mu_0,
\phi)$.  Thus, we can restrict to $\mu = 2\mu_0 + 4k\mu_0, k \ge 0$
where the equation {\em is} deterministic.  

In both cases then, the space of solutions of the constraint equation,
is completely determined by giving appropriate data for large $|\mu|$
i.e. in the classical regime. Such a deterministic nature of the
constraint equation has been taken as a necessary condition for
non-singularity at the quantum level \footnote{For contrast, if one just
symmetrizes the non-symmetric operator (without the sgn factor), one
gets a difference equation which {\em is non-deterministic}.}. As such
this could be viewed as a criterion to limit the choice of factor
ordering.

By introducing an interpolating, slowly varying smooth function, $\Psi(p
(\mu) := \Case{1}{6}\gamma\lP^2)$, and keeping only the first
non-vanishing terms, one deduces the Wheeler-De Witt differential
equation (with a modified matter Hamiltonian) from the above difference
equation. Making a WKB approximation, one infers an effective
Hamiltonian which matches with the classical Hamiltonian for large
volume ($\mu \gg \mu_0$) and small extrinsic curvature (derivative of
the WKB phase is small). There are terms of $o(\hbar^0)$ which contain
arbitrary powers of the first derivative of the phase which can all be
summed up. The resulting effective Hamiltonian now contains
modifications of the classical gravitational Hamiltonian, apart from the
modifications in the matter Hamiltonian due to the inverse powers of the
triad. The largest possible domain of validity of effective Hamiltonian
so deduced must have $|p| \gtrsim p_0$ \cite{SemiClass,EffHam}.

An effective Hamiltonian can alternatively obtained by computing
expectation values of the Hamiltonian operator in semiclassical states
peaked in classical regimes \cite{Willis}. The leading order effective
Hamiltonian that one obtains is (spatially flat case):
\begin{eqnarray}
H^{\mathrm{non-sym}}_{\mathrm{eff}} & = & - \frac{1}{16 \pi
G}\left(\frac{6}{\mu_0^3 \gamma^3 \lP^2}\right) \left[ B_+(p)
\mathrm{sin}^2(\mu_0c) + \left( A(p) - \frac{1}{2}B_+(p) \right) \right]
+ H_{\mathrm{matter}} \ ; \nonumber\\
B_+(p) & := & A(p + 4 p_0) + A(p - 4 p_0) ~,~ A(p) ~ := ~ (|p +
p_0|^{3/2} - |p - p_0|^{3/2}) \ , \\
p & := & \frac{1}{6}\gamma \lP^2 \mu ~ ~ , ~ ~ p_0 ~ := ~
\frac{1}{6}\gamma \lP^2 \mu_0 \ . \nonumber
\end{eqnarray}
For the symmetric operator, the effective Hamiltonian is the same as
above except that $B_+(p) \to f_+(p) + f_-(p)$ and $2 A(p) \to f_+(p) +
f_-(p)$. 

The second bracket in the square bracket, is the quantum geometry
potential which is negative and higher order in $\lP$ but is important
in the small volume regime and plays a role in the genericness of bounce
deduced from the effective Hamiltonian \cite{GenBounce}. This term is
absent in effective Hamiltonian deduced from the symmetric constraint.
The matter Hamiltonian will typically have the eigenvalues of powers of
inverse triad operator which depend on the ambiguity parameters $j, l$.

We already see that the quantum modifications are such that both the
matter and the gravitational parts in the effective Hamiltonian, are
rendered bounded and effective dynamics must be non-singular.

For large values of the triad, $p \gg p_0$, $B_+(p) \sim 6 p_0 \sqrt{p}
- o(p^{-3/2})$ while $A(p) \sim 3 p_0 \sqrt{p} - o(p^{-3/2})$. In this
regime, the effective Hamiltonians deduced from both symmetric and
non-symmetric ordering are the same. The classical Hamiltonian is
obtained for $\mu_0 \to 0$. From this, one can obtain the equations of
motion and by computing the left hand side of the Friedmann equation,
infer the effective energy density. For $p \gg p_0$ one obtains
\footnote{For $p$ in the semiclassical regime, one should include the
contribution of the quantum geometry potential present in the
non-symmetric ordering, especially for examining the bounce
possibility \cite{EffHam}.},
\begin{equation} \label{EffHam}
\frac{3}{8\pi G}\left(\frac{\dot{a}^2}{a^2}\right) ~ := ~
\rho_{\mathrm{eff}} ~ = ~
\left(\frac{H_{\mathrm{matter}}}{p^{3/2}}\right)\left\{1 - \frac{8 \pi G
\mu_0^2\gamma^2}{3} p
\left(\frac{H_{\mathrm{matter}}}{p^{3/2}}\right)\right\} ~ ~,~ ~ p :=
a^2/4\ .
\end{equation}

The effective density is quadratic in the classical density, $\rho_{cl}
:= H_\mathrm{matter} p^{-3/2}$. This modification is due to the quantum
correction in the gravitational Hamiltonian (due to the sin$^2$
feature). This is over and above the corrections hidden in the matter
Hamiltonian (due to the ``inverse volume'' modifications). As noted
before, we have two scales: $p_0$ controlled by $\mu_0$ in the
gravitational part and $2p_0 j$ in the matter part. For large $j$ it is
possible that we can have $p_0 \ll p \ll 2p_0j $ in which case the above
expressions will hold with $j$ dependent corrections in the matter
Hamiltonian. In this semiclassical regime, the corrections from sin$^2$
term are smaller in comparison to those from inverse volume. If $p \gg
2p_0j$ then the matter Hamiltonian is also the classical expression. For
$j = 1/2$, there is only the $p \gg p_0$ regime and $\rho_{cl}$ is
genuinely the classical density.

Let us quickly note a comparison of the two quantizations as reflected
in the corresponding effective Hamiltonians, particularly with regards
to the extrema of $p(t)$. For this, we will assume same ambiguity
parameters $(j, l)$ in the matter Hamiltonian, $(1/2) p_{\phi}^2
F_{jl}^{3/2}$ and explore the regime $p \gtrsim p_0$. The two effective
Hamiltonians differ significantly in the semiclassical regime due to the
quantum geometry potential.

The equations of motion imply that $p_{\phi}$ is a constant of motion,
$\phi$ varies monotonically with $t$ and on the constraint surface, we
can eliminate $c$ in favour of $p$ and $p_{\phi}$. Let us focus on
$p(t)$ and in particular consider its possible extrema.  It is immediate
that $\dot{p} = 0$ implies sin($2\mu_0c$) = 0. This leads to two
possibilities: (A) sin($\mu_0c$) = 0 or (B) cos($\mu_0c$) = 0. A local
{\em minimum} signifies a {\em bounce} while a local {\em maximum}
signifies a {\em re-collapse}. The value of $p$ at an extremum, $p_*$,
gets determined in terms of the constant $p_{\phi}$. The
bounce/re-collapse nature of an extremum depends upon whether $p_*$ is
in the classical regime or in the semiclassical regime and also on the
case (A) or (B). Note that for the case (A) to hold, it is necessary
that the quantum geometry potential is present. Thus, for the symmetric
ordering, case (A) cannot be realised -- it will imply $p_{\phi} = 0$.

An extremum determined by case (A): It is a bounce if $p_*$ is in in the
{\em semiclassical} regime; $p_*$ varies inversely with $p_{\phi}$ while
the corresponding density varies directly. $p_*$ being limited to the
semiclassical regime implies that $p_{\phi}$ is also bounded both above
and below, for such an extremum to occur. It turns out that $p_*$ can be
in the {\em classical} regime, provided $p_{\phi} \sim \lP^2$. Thus, the
non-symmetric constraint, at the effective level, can accommodate a
bounce only in the semiclassical regime and with large densities. 

An extremum determined by case (B): It is bounce if $p_*$ is in the {\em
classical} regime; $p_*$ varies directly with $p_{\phi}$ and the
corresponding density varies inversely. $p_*$ being limited to the
classical regime implies that $p_{\phi}$ must be bounded below but can
be arbitrarily large and thus the density can be arbitrarily small. This
is quite unreasonable and has been sited as one of the reasons for
considering the ``improved'' quantization (more on this later). If $p_*$
is in the {\em semiclassical} regime, it has to be a re-collapse with
$p_{\phi} \sim \lP^2$.

In the early works, one worked with the non-symmetric constraint
operator and the sin$^2$ corrections were not incorporated (i.e.
$\mu_0c \ll \pi/2$ was assumed) and the phenomenological implications
were entirely due to the modified matter Hamiltonian. These already
implied genericness of inflation and genericness of bounce. These
results were discussed at the previous IAGRG meeting in Jaipur. 

To summarize: LQC differs from the earlier quantum cosmology in three
basic ways (a) the basic variables are different and in particular the
classical singularity is in the interior of the mini-superspace; (b) the
quantization is very different, being motivated by the background
independent quantization employed in LQG; (c) there is a ``parent''
quantum theory (LQG) which is pretty much well defined at the
kinematical level, unlike the metric variables based Wheeler-De Witt
theory. The loop quantization has fundamentally distinct implications:
its discrete nature of quantum geometry leads to bounded energy
densities and bounded extrinsic and intrinsic curvatures (for the
anisotropic models). These two features are construed as ``resolving the
classical singularity''.  Quite un-expectedly, the effective dynamics
incorporating quantum corrections is also singularity-free (via a
bounce), accommodates an inflationary phase rather naturally and is well
behaved with regards to perturbations. Although there are many ambiguity
parameters, these results are robust with respect to their values.

\section{Post 2004 Isotropic LQC}

Despite many attractive features of LQC, many points need to be
addressed further:
\begin{itemize}
\item LQC being a constrained theory, it would be more appropriate if
singularity resolution is formulated and demonstrated in terms of
physical expectation values of physical (Dirac) operators i.e. in terms
of ``gauge invariant quantities''. This can be done at present with
self-adjoint constraint i.e. a symmetric ordering and for free, massless
scalar matter.
\item There are at least three distinct ambiguity parameters: $\mu_0$
related to the fiducial length of the loop used in writing the
holonomies; $j$ entering in the choice of $SU(2)$ representation which
is chosen to be $1/2$ in the gravitational sector and some large value
in the matter sector; $l$ entering in writing the inverse powers in
terms of Poisson brackets. The first one was thought to be determined by
the area gap from the full theory. The $j = 1/2$ in the gravitational
Hamiltonian seems needed to avoid high order difference equation and
larger $j$ values are hinted to be problematic in the study of a three
dimensional model \cite{LargeJ}.  Given this, the choice of a high value
of $j$ in the matter Hamiltonian seems unnatural\footnote{For an
alternative view on using large values of $j$, see reference
\cite{MartinLattice}.}. For phenomenology however the higher values
allowing for a larger semiclassical regime are preferred. The $l$ does
not play as significant a role.
\item The bounce scale and density at the bounce, implied by the
effective Hamiltonian (from symmetric ordering), is dependent on the
parameters of the matter Hamiltonian and can be arranged such that the
bounce density is arbitrarily small.  This is a highly undesirable
feature. Furthermore, the largest possible domain of validity of WKB
approximation is given by the turning points (eg the bounce scale).
However, the approximation could break down even before reaching the
turning point. An independent check on the domain of validity of
effective Hamiltonian is thus desirable.
\item A systematic derivation of LQC from LQG is expected to tighten the
ambiguity parameters. However, such a derivation is not yet available.
\end{itemize}

\subsection{Physical quantities and Singularity Resolution}

When the Hamiltonian is a constraint, at the classical level itself, the
notion of dynamics in terms of the `time translations' generated by the
Hamiltonian is devoid of any physical meaning.  Furthermore, at the
quantum level when one attempts to impose the constraint as
$\hat{H}|\Psi\rangle = 0$, typically one finds that there are no
solutions in the Hilbert space on which $\hat{H}$ is defined - the
solutions are generically distributional. One then has to consider the
space of all distributional solutions, define a new physical inner
product to turn it into a Hilbert space (the physical Hilbert space),
define operators on the space of solutions (which must thus act
invariantly) which are self-adjoint (physical operators) and compute
expectation values, uncertainties etc of these operators to make
physical predictions.  Clearly, the space of solutions depends on the
quantization of the constraint and there is an arbitrariness in the
choice of physical inner product. This is usually chosen so that a
complete set of Dirac observables (as deduced from the classical theory)
are self-adjoint. This is greatly simplified if the constraint has a
{\em separable} form with respect to some degree of freedom \footnote{A
general abstract procedure using group averaging is also available.  }.
For LQC (and also for the Wheeler-De Witt quantum cosmology), such a
simplification is available for a free, massless scalar matter:
$H_{\mathrm{matter}}(\phi, p_{\phi}) := \Case{1}{2} p_{\phi}^2
|p|^{-3/2}$. Let us sketch the steps schematically, focusing on the
spatially flat model for simplicity \cite{APSOne,APSTwo}.
\begin{enumerate}
\item {\em Fundamental constraint equation:} 

The classical constraint equations is:
\begin{equation}
 - \frac{6}{\gamma^2} c^2 \sqrt{|p|} + 8\pi G \ p_{\phi}^2 \ |p|^{-3/2}
 ~ = ~ 0 ~ = ~ C_{\mathrm{grav}} + C_{\mathrm{matter}}\ ;
\end{equation}
The corresponding quantum equation for the wave function, $\Psi(p,
\phi)$ is:
\begin{equation}
8\pi G \hat{p}_{\phi}^2 \Psi(p, \phi) ~ = ~ [\tilde{B}(p)]^{-1}
\hat{C}_{\mathrm{grav}} \Psi(p, \phi)~ ~ , ~ ~ [\tilde{B}(p)] \mbox{ is
eigenvalue of } \widehat{|p|^{-3/2}} \ ;
\end{equation}
Putting $\hat{p}_{\phi} = - i \hbar \partial_{\phi}$, $p :=
\Case{\gamma\lP^2}{6}\mu$  and $\tilde{B}(p) :=
(\Case{\gamma\lP^2}{6})^{-3/2} B(\mu)$, the equation can be written in a
separated form as, 
\begin{equation} 
\frac{\partial^2 \Psi(\mu, \phi)}{\partial \phi^2} ~ = ~
[B(\mu)]^{-1}\left[8\pi G \left(\frac{\gamma}{6}\right)^{3/2}\lP^{-1}
\hat{C}_{\mathrm{grav}}\right]\Psi(\mu, \phi) ~ := ~ -
\hat{\Theta}(\mu)\Psi(\mu, \phi).
\end{equation}
The $\hat{\Theta}$ operator for different quantizations is different.
For Schrodinger quantization (Wheeler-De Witt), with a particular factor
ordering suggested by the continuum limit of the difference equation,
the operator $\hat{\Theta}(\mu)$ is given by,
\begin{equation}
\hat{\Theta}_{\mathrm{Sch}}(\mu)\Psi(\mu, \phi) ~ = ~ - \frac{16\pi
G}{3}|\mu|^{3/2} \partial_{\mu} \sqrt{\mu}\ \partial_{\mu}\Psi(\mu,
\phi)
\end{equation}
while for LQC, with symmetric ordering, it is given by,
\begin{eqnarray}
\hat{\Theta}_{\mathrm{LQC}}(\mu)\Psi(\mu, \phi) & = & - [B(\mu)]^{-1}
\left\{ C^+(\mu) \Psi(\mu + 4 \mu_0, \phi) + C^0(\mu) \Psi(\mu, \phi) +
\right. \nonumber \\ 
& & \left. \hspace{5.0cm} C^-(\mu) \Psi(\mu - 4 \mu_0, \phi) \right\} \
, \nonumber \\ 
C^+(\mu) & := & \frac{\pi G}{9 \mu_0^3} \left| ~ |\mu + 3\mu_0|^{3/2} -
|\mu + \mu_0|^{3/2} \right| \ , \\
C^-(\mu) & := & C^+(\mu - 4 \mu_0) ~ ~ , ~ ~ C^0(\mu) ~ := ~ - C^+(\mu)
- C^-(\mu) \ . \nonumber
\end{eqnarray}
Note that in the Schrodinger quantization, the $B_{\mathrm{Sch}}(\mu) =
|\mu|^{-3/2}$ diverges at $\mu = 0$ while in LQC,
$B_{\mathrm{LQC}}(\mu)$ vanishes for all allowed choices of ambiguity
parameters. In both cases, $B(\mu) \sim |\mu|^{-3/2}$ as $|\mu| \to
\infty$.

\item {\em Inner product and General solution:} 

The operator $\hat{\Theta}$ turns out to be a self-adjoint, positive
definite operator on the space of functions $\Psi(\mu, \phi)$ for each
fixed $\phi$ with an inner product scaled by $B(\mu)$. That is, for the
Schrodinger quantization, it is an operator on $L^2(\mathbb{R},
B_{\mathrm{Sch}}(\mu) d\mu)$ while for LQC it is an operator on
$L^2(\mathbb{R}_{\mathrm{Bohr}},
B_{\mathrm{Bohr}}(\mu)d\mu_{\mathrm{Bohr}}).$ Because of this, the
operator has a complete set of eigenvectors: $\hat{\Theta}e_k(\mu) =
\omega^2(k) e_k(\mu), k \in \mathbb{R}, \langle e_k|e_{k'}\rangle =
\delta(k, k')$, and the general solution of the fundamental constraint
equation can be expressed as
\begin{equation}
\Psi(\mu, \phi) ~ = ~ \int dk ~ \tilde{\Psi}_+(k) e_k(\mu)
e^{i\omega\phi} + \tilde{\Psi}_-(k) \bar{e}_k(\mu)  e^{-i \omega\phi} \
.
\end{equation}
The orthonormality relations among the $e_k(\mu)$ are in the
corresponding Hilbert spaces. Different quantizations differ in the form
of the eigenfunctions, possibly the spectrum itself and of course
$\omega(k)$. In general, these solutions are {\em not} normalizable in
$L^2(\mathbb{R}_{\mathrm{Bohr}}\times\mathbb{R},
d\mu_{\mathrm{Bohr}}\times d\mu)$, i.e. these are distributional.

\item {\em Choice of Dirac observables:}

Since the classical kinematical phase space is 4 dimensional and we have
a single first class constraint, the phase space of physical states
(reduced phase space) is two dimensional and we need two functions to
coordinatize this space. We should thus look for two (classical) Dirac 
observables: functions on the kinematical phase space whose Poisson
bracket with the Hamiltonian constraint vanishes on the constraint
surface. 

It is easy to see that $p_{\phi}$ is a Dirac observable. For the second
one, we choose a one parameter family of functions $\mu(\phi)$
satisfying $\{\mu(\phi), C(\mu, c, \phi, p_{\phi}) \} \approx 0$. The
corresponding quantum definitions, with the operators acting on the
solutions, are:
\begin{eqnarray}
\hat{p}_{\phi}\Psi(\mu, \phi) & := & -i \hbar\partial_{\phi}\Psi(\mu,
\phi) \ , \\
\widehat{|\mu|_{\phi_0}}\Psi(\mu, \phi) & := &
e^{i\sqrt{\hat{\Theta}}(\phi - \phi_0)} |\mu| \Psi_+(\mu, \phi_0) + e^{-
i\sqrt{\hat{\Theta}}(\phi - \phi_0)} |\mu| \Psi_-(\mu, \phi_0)
\end{eqnarray}
On an initial datum, $\Psi(\mu, \phi_0)$, these operators act as,
\begin{equation}
\widehat{|\mu|_{\phi_0}}\Psi(\mu, \phi_0) ~ = ~ |\mu|\Psi(\mu, \phi_0) ~
~ , ~ ~ \hat{p}_{\phi} \Psi(\mu, \phi_0) ~ = ~ \hbar
\sqrt{\hat{\Theta}}\Psi(\mu, \phi_0)\ .
\end{equation}

\item {\em Physical inner product:}

It follows that the Dirac operators defined on the space of solutions
are self-adjoint if we define a {\em physical inner product} on the
space of solutions as:
\begin{equation}
\langle\Psi|\Psi'\rangle_{\mathrm{phys}} ~ := ~ ``\int_{\phi = \phi_0}
d\mu B(\mu)" ~ ~ \bar{\Psi}(\mu, \phi) \Psi'(\mu, \phi) \ .
\end{equation}
Thus the eigenvalues of the inverse volume operator crucially enter the
definition of the physical inner product. For Schrodinger quantization,
the integral is really an integral while for LQC it is actually a sum
over $\mu$ taking values in a lattice. The inner product is independent
of the choice of $\phi_0$.

A complete set of physical operators and physical inner product has now
been specified and physical questions can be phrased in terms of
(physical) expectation values of functions of these operators.

\item {\em Semiclassical states:}

To discuss semiclassical regime, typically one {\em defines}
semiclassical states: physical states such that a chosen set of
self-adjoint operators have specified expectation values with
uncertainties bounded by specified tolerances. A natural choice of
operators for us are the two Dirac operators defined above. It is easy
to construct semiclassical states with respect to these operators. For
example, a state peaked around, $p_{\phi} =  p^*_{\phi}$ and
$|\mu|_{\phi_0} = \mu^*$ is given by (in Schrodinger quantization for
instance),
\begin{eqnarray}
\Psi_{\mathrm{semi}}(\mu, \phi_0) & := & \int dk e^{- \frac{(k -
k^*)^2}{2 \sigma^2}} e_k(\mu) e^{i \omega (\phi_0 - \phi^*)} \\ k^* & =
& -\sqrt{3/2\kappa} \hbar^{-1} p^*_{\phi} ~ ~ ,  ~ ~ \phi^* = \phi_0 +
-\sqrt{3/2\kappa} \ell n|\mu^*| \ .
\end{eqnarray}
For LQC, the $e_k(\mu)$ functions are different and the physical
expectation values are to be evaluated using the physical inner product
defined in the LQC context.

\item {\em Evolution of physical quantities:}

Since one knows the general solution of the constraint equation,
$\Psi(\mu, \phi)$, given $\Psi(\mu, \phi_0)$, one can compute the
physical expectation values in the semiclassical solution,
$\Psi_{\mathrm{semi}}(\mu, \phi)$ and track the position of the peak as
a function of $\phi$ as well as the uncertainties as a function of
$\phi$. 

\item {\em Resolution of Big Bang Singularity:}

A classical solution is obtained as a curve in $(\mu, \phi)$ plane,
different curves being labelled by the points ($\mu^*, \phi^*$) in the
plane. The curves are independent of the constant value of
$p^{*}_{\phi}$ These curves are already given in
(\ref{ClassRelationalSoln}).

Quantum mechanically, we first select a semiclassical solution,
$\Psi_{\mathrm{semi}}(p_{\phi}^*, \mu^*: \phi)$ in which the expectation
values of the Dirac operators, at $\phi = \phi_0$, are $p_{\phi}^*$ and
$\mu^*$ respectively. These values serve as labels for the semiclassical
solution. The former one continues to be $p_{\phi}^*$ for all $\phi$
whereas $\langle\widehat{|\mu|_{\phi_0}}\rangle(\phi) =:
|\mu|_{p_{\phi}^*, \mu^*}(\phi)$, determines a curve in the $(\mu,
\phi)$ plane. In general one expects this curve to be different from the
classical curve in the region of small $\mu$ (small volume).

The result of the computations is that Schrodinger quantization, the
curve $ |\mu|_{p_{\phi}^*, \mu^*}(\phi)$, does approach the $\mu = 0$
axis asymptotically. However for LQC, the curve {\em bounces away} from
the $\mu = 0$ axis. In this sense -- and now inferred in terms of
physical quantities -- the Big Bang singularity is resolved in LQC.  It
also turns out that for large enough values of $p^*_{\phi}$, the quantum
trajectories constructed by the above procedure are well approximated by
the trajectories by the effective Hamiltonian. All these statements are
for semiclassical solutions which are peaked at large $\mu_*$ at late
times.
\end{enumerate}

Two further features are noteworthy as they corroborate the suggestions
from the effective Hamiltonian analysis. 

First one is revealed by computing expectation value of the matter
density operator, $\rho_{\mathrm{matter}} :=
\Case{1}{2}\widehat{(p_{\phi}^*)^2 |p|^{-3}}$, at the bounce value of
$|p|$. It turns out that this value is sensitive to the value of
$p_{\phi}^*$ and can be made arbitrarily {\em small} by choosing
$p_{\phi}^*$ to be {\em large}. Physically this is unsatisfactory as
quantum effects are {\em not} expected to be significant for matter
density very small compared to the Planck density. This is traced to the
quantization of the gravitational Hamiltonian, in particular to the step
which introduces the ambiguity parameter $\mu_0$. A novel solution
proposed in the ``improved quantization'', removes this undesirable
feature. 

The second one refers to the role of quantum modifications in the
gravitational Hamiltonian compared to those in the matter Hamiltonian
(the inverse volume modification or $B(\mu)$). The former is much more
significant than the latter. So much so, that even if one uses the
$B(\mu)$ from the Schrodinger quantization (i.e. switch-off the inverse
volume modifications), one still obtains the bounce.  So bounce is seen
as the consequence of $\hat{\Theta}$ being different and as far as
qualitative singularity resolution is concerned, the inverse volume
modifications are {\em un-important}. As the effective picture (for
symmetric constraint) showed, the bounce occurs in the classical region
(for $j = 1/2$) where the inverse volume corrections can be neglected.
For an exact model which seeks to understand as to why the bounces are
seen, please see \cite{MartinExact}.

\subsection{Improved Quantization}
The undesirable features of the bounce coming from the classical region,
can be seen readily using the effective Hamiltonian, as remarked
earlier. To see the effects of modifications from the gravitational
Hamiltonian, choose $j = 1/2$ and consider the Friedmann equation
derived from the effective Hamiltonian (\ref{EffHam}), with matter
Hamiltonian given by $H_{\mathrm{matter}} = \Case{1}{2}p^2_{\phi}
|p|^{-3/2}$. The positivity of the effective density implies that $p \ge
p_*$ with $p_*$ determined by vanishing of the effective energy density:
$\rho_* := \rho_{cl}(p_*) = (\Case{8\pi G \mu_0^2 \gamma^2}{3}
p_*)^{-1}$. This leads to $|p_*| = \sqrt{\Case{4 \pi G \mu_0^2
\gamma^2}{3}} |p_{\phi}|$ and $\rho_* = \sqrt{2} (\frac{8\pi G \mu_0^2
\gamma^2}{3})^{-3/2} |p_{\phi}|^{-1}$.  One sees that for large
$|p_{\phi}|$, the bounce scale $|p_*|$ can be large and the maximum
density -- density at bounce -- could be small.  Thus, {\em within the
model}, there exist a possibility of seeing quantum effects (bounce)
even when neither the energy density nor the bounce scale are comparable
to the corresponding Planck quantities and this is an undesirable
feature of the model. This feature is independent of factor
ordering as long as the bounce occurs in the classical regime. 

One may notice that {\em if} we replace $\mu_0 \to \bar{\mu}(p) :=
\sqrt{\Delta/|p|}$ where $\Delta$ is a constant, then the effective
density vanishes when $\rho_{cl}$ equals the critical value
$\rho_{\mathrm{crit}} := (\Case{8\pi G \Delta \gamma^2}{3})^{-1}$, which
is independent of matter Hamiltonian. The bounce scale $p_*$ is
determined by $\rho_* = \rho_{\mathrm{crit}}$ which gives $|p_*| =
(\Case{p_{\phi}^2}{2\rho_{\mathrm{crit}}})^{1/3}$. Now although the
bounce scale can again be large depending upon $p_{\phi}$, the density
at bounce is always the universal value determined by $\Delta$. This is
a rather nice feature in that quantum geometry effects are revealed when
matter density (which couples to gravity) reaches a universal, critical
value regardless of the dynamical variables describing matter. For a
suitable choice of $\Delta$ one can ensure that a bounce always happen
when {\em the energy density} becomes comparable to the Planck density.
In this manner, one can retain the good feature (bounce) even for $j =
1/2$ thus ``effectively fixing'' an ambiguity parameter and also trade
another ambiguity parameter $\mu_0$ for $\Delta$. This is precisely what
is achieved by the ``improved quantization'' of the gravitational
Hamiltonian \cite{APSThree}.

The place where the quantization procedure is modified is when one
expresses the curvature in terms of the holonomies along a loop around a
``plaquette''. One shrinks the plaquette in the limiting procedure.  One
now makes an important departure: the plaquette should be shrunk only
till the physical area (as distinct from a fiducial one) reaches its
minimum possible value which is given by the area gap in the known
spectrum of area operator in quantum geometry: $\Delta = 2\sqrt{3}\pi
\gamma G\hbar$. Since the plaquette is a square of fiducial length
$\mu_0$, its physical area is $\mu_0^2|p|$ and this should set be to
$\Delta$. Since $|p|$ is a dynamical variable, $\mu_0$ cannot be a
constant and is to be thought of a function on the phase space,
$\bar{\mu}(p) := \sqrt{\Delta/|p|}$. It turns out that even with such a
change which makes the curvature to be a function of both connection and
triad, the form of {\em both} the gravitational constraint and inverse
volume operator appearing in the matter Hamiltonian, remains the same
with just doing the replacement, $\mu_0 \to \bar{\mu}$ defined above,
{\em in the holonomies}.  The expressions simplify by using
eigenfunctions of the volume operator $\hat{V} := \hat{|p|}^{3/2}$,
instead of those of the triad. The relevant expressions are:
\begin{eqnarray}
v & := & K \mathrm{sgn}(\mu) |\mu|^{3/2}~ ~,~ ~ K ~ := ~ \frac{2
\sqrt{2}}{3 \sqrt{3\sqrt{3}}}~; \\
\hat{V}|v\rangle & = &
\left(\frac{\gamma}{6}\right)^{3/2}\frac{\lP^3}{K}|v| |v\rangle~,~ \\
\widehat{e^{ik\Case{\bar{\mu}}{2}c}}\Psi(v) & := & \Psi(v + k)~, \\
\left.\widehat{|p|^{-1/2}}\right|_{j = 1/2, l = 3/4}\Psi(v) & = &
\frac{3}{2}\left(\frac{\gamma\lP^2}{6}\right)^{-1/2}
K^{1/3}|v|^{1/3}\left| |v + 1|^{1/3} - |v - 1|^{1/3}\right| \Psi(v) \\
B(v) & = & \left(\frac{3}{2}\right)^{3/2} K |v| \left| |v + 1|^{1/3} -
|v - 1|^{1/3}\right|^3 \\
\hat{\Theta}_{\mathrm{Improved}}\Psi(v, \phi) & = & - [B(v)]^{-1}
\left\{ C^+(v) \Psi(v + 4, \phi) + C^0(v) \Psi(v, \phi) + \right.
\nonumber \\ 
& & \left. \hspace{5.0cm} C^-(v) \Psi(v - 4, \phi) \right\} \ , \\ 
C^+(v) & := & \frac{3 \pi K G}{8} |v + 2| \left| ~ |v + 1| - |v + 3|
\right| \ , \\
C^-(v) & := & C^+(v - 4) ~ ~ , ~ ~ C^0(v) ~ := ~ - C^+(v) - C^-(v) \ . 
\end{eqnarray}

Thus the main changes in the quantization of the Hamiltonian constraint
are: (1) replace $\mu_0 \to \bar{\mu} := \sqrt{\Delta/|p|}$ in the
holonomies; (2) {\em choose} symmetric ordering for the gravitational
constraint; and (3) {\em choose} $j = 1/2$ in both gravitational
Hamiltonian and the matter Hamiltonian (in the definition of inverse
powers of triad operator). The ``improvement'' refers to the first
point. This model is singularity free at the level of the fundamental
constraint equation (even though the leading coefficients of the
difference equation do vanish, because the the parity symmetry again
saves the day); the densities continue to be bounded above -- and now
with a bound independent of matter parameters; the effective picture
continues to be singularity free and with undesirable features removed
and the classical Big Bang being replaced by a quantum bounce is
established in terms of {\em physical} quantities.

\subsection{Close Isotropic Model}

While close model seems phenomenologically disfavoured, it provides
further testing ground for quantization of the Hamiltonian constraint.
Because of the intrinsic (spatial) curvature, the plaquettes used in
expressing the $F_{ij}$ in terms of holonomies, are not bounded by just
four edges -- a fifth one is necessary. This was attempted and was found
to lead to an ``unstable'' quantization. This difficulty was bypassed by
using the holonomies of the extrinsic curvature instead of the gauge
connection which is permissible in the homogeneous context. The
corresponding, non-symmetric constraint and its difference equation was
analysed for the massless scalar matter. Green and Unruh, found that
solutions of the difference equation was always diverging (at least for
one orientation) for large volumes. Further, the divergence seemed to
set in just where one expected a re-collapse from the classical theory.
In the absence of physical inner product and physical interpretation of
the solutions, it was concluded that this version of LQC for close model
is unlikely to accommodate classical re-collapse even though it avoided
the Big Bang/Big Crunch singularities. 

Recently, this model has been revisited \cite{APSFour}. One went back to
using the gauge connection and the fifth edge difficulty was
circumvented by using both the left-invariant and the right-invariant
vector fields to define the plaquette. In addition, the symmetric
ordering was chosen and finally the $\mu_0 \to \bar{\mu}$ improvement
was also incorporated.  Without the improvement, there were still the
problems of getting bounce for low energy density and also not getting a
reasonable re-collapse (either re-collapse is absent or the scale is
marginally larger than the bounce scale). With the improvement, the
bounces and re-collapses are neatly accommodated and one gets a cyclic
evolution. In this case also, the scalar field serves as a good clock
variable as it continues to be monotonic with the synchronous time.

I have focussed on the singularity resolution issue in this talk.
Other developments have also taken place in the past couple of
years. I will just list these giving references.

\begin{enumerate}
\item {\em Effective models and their properties:} The effective picture
was shown to be non-singular and since this is based on the usual
framework of GR, it follows that energy conditions must be violated (and
indeed they are thanks to the inverse volume modifications). This raised
questions regarding stability of matter and causal propagation of
perturbations. Golam Hossain showed that despite the energy conditions
violations, neither of the above pathologies result \cite{EnergyCondns}. 

Minimally coupled scalar has been used in elaborating inflationary
scenarios. However non-minimally coupled scalars are also conceivable
models. The singularity resolution and inflationary scenarios continue
to hold also in this case. Furthermore sufficient e-foldings are also
admissible \cite{NonMinimal}.

In the improved quantization, one sets the ambiguity parameter $j = 1/2$
and shifts the dominant effects to the the gravitational Hamiltonian.
All the previous phenomenological implications however were driven by
the inverse volume modifications in the matter sector.  Consequently, it
is necessary to check if and how the phenomenology works with the
improved quantization. This has been explored in
\cite{ImprovedPhenomenology}.

Using the effective dynamics for the homogeneous mode, density
perturbations were explored and power spectra were computed with
the required small amplitude \cite{DensityPert,Calcagni}. 

As many of the phenomenology oriented questions have been explored using
effective Hamiltonian which incorporate quantum corrections from various
sources (gravity, matter etc). This motivates a some what systematic
approach to constructing effective approximations. This has been
initiated in \cite{EffectiveTheories}.

\item {\em Anisotropic models:} The anisotropic models provide further
testing grounds for loop quantization. At the difference equation level,
the non-singularity has been checked also for these models in the
non-symmetric scheme. For the vacuum Bianchi I model,
there is no place for the inverse volume type corrections to appear at
an effective Hamiltonian level and the effective dynamics would continue
to be singular. However, once the gravitational corrections (sin$^2$)
are incorporated, the effective dynamics again is non-singular and one
can obtain the non-singular version of the (singular) Kasner solution
\cite{KasnerDate}. More recently, the Bianchi I model with a free,
massless scalar is also analysed in the improved quantization
\cite{KasnerScalar}. A perturbative treatment of anisotropies has been
explored in \cite{PerturbAniso}.

\item {\em Inhomogeneities:} Inhomogeneities are a fact of nature
although these are small in the early universe. This suggests a
perturbative approach to incorporate inhomogeneities. On the one hand
one can study their evolution in the homogeneous, isotropic background
(cosmological perturbation theory). One can also begin with a
(simplified) inhomogeneous model and try to see how a homogeneous
approximation can become viable. The work on the former has already
begun. For the latter part, Bojowald has discussed a simplified lattice
model to draw some lessons for the homogeneous models. In particular he
has given an alternative argument for the $\mu_0 \to \bar{\mu}$
modification which does not appeal to the area operator
\cite{MartinLattice}.
\end{enumerate}

\section{Open Issues and Out look}

In summary, over the past two years, we have seen how to phrase and
understand the fate of Big Bang singularity in a quantum framework.

Firstly, with the help of a minimally coupled, free, massless scalar
which serves as a good clock variable in the isotropic context, one can
define physical inner product, a complete set of Dirac observables and
their physical matrix elements. At present this can be done only for
self-adjoint Hamiltonian constraint. Using these, one can construct
trajectories in the $(p, p_{\phi})$ plane which are followed by the peak
of a semiclassical state as well as the uncertainties in the Dirac
observables.  It so happens that these trajectories do not pass through
the zero volume -- Big Bang is replaced by a Bounce. For close isotropic
model, the Big Crunch is also replaced by a bounce while retaining
classically understood re-collapse. In conjunction with the $\bar{\mu}$
improvement, the gravitational Hamiltonian can be given the the main
role in generating the bounce. A corresponding treatment in Schrodinger
quantization (Wheeler-De Witt theory), {\em does not} generate a bounce
nor does it render the density, curvatures bounded. Thus, quantum
representation plays a significant role in the singularity resolution. 

Secondly, the improved quantization motivated by the regulation of the
$F_{ij}$ invoking the area operator from the full theory (or by the
argument from the inhomogeneous lattice model), also leads the bounce to
be ``triggered'' when the energy density reaches a critical value ($\sim
0.82 \rho_{\mathrm{Planck}}$) which is {\em independent} of the values
of the dynamical variables.  Close model also gives the same critical
value.

While the improvement is demonstrated to be viable in the isotropic
context, the procedure differs from that followed in the full theory.
One may either view this as something special to the mini-superspace
model(s) or view it as providing hints for newer approaches in the full
theory.

A general criteria for ``non-singularity'' is not in sight yet and so
also a systematic derivation of the mini-superspace model(s) from a
larger, full theory.

{\em Acknowledgements:} I would like to thank Parampreet Singh for
discussions regarding the improved quantization as well as for running
his codes for exploring bounce in the semiclassical regime. Thanks are
due to Martin Bojowald for comments on an earlier draft.

\end{document}